\documentclass[prodmode,acmtap]{ci2014}

\acmVolume{2}
\acmNumber{3}
\acmArticle{1}
\articleSeq{1}
\acmYear{2013}
\acmMonth{5}

\usepackage[ruled]{algorithm2e}
\usepackage{multirow}
\usepackage{booktabs}
\usepackage{amsmath}
\usepackage{array}
\newcolumntype{P}[1]{>{\centering\arraybackslash}p{#1}}
\newcommand{\ra}[1]{\renewcommand{\arraystretch}{#1}}
\usepackage{caption}
\usepackage{caption}
\SetAlFnt{\algofont}
\SetAlCapFnt{\algofont}
\SetAlCapNameFnt{\algofont}
\SetAlCapHSkip{0pt}
\IncMargin{-\parindent}
\renewcommand{\algorithmcfname}{ALGORITHM}

\title{Nowcasting the Bitcoin Market with Twitter Signals\footnote{Last Update: September 2014. An early version of this paper was presented at Collective Intelligence 2014, MIT, Cambridge, USA, June 10-12 2014.}}
\author{JERMAIN C. KAMINSKI, MIT Media Lab{}
}

\begin{abstract}
This paper analyzes correlations and causalities between Bitcoin market indicators and Twitter posts containing emotional signals on Bitcoin. Within a timeframe of 104 days  (November 23$^{rd}$ 2013 - March 7$^{th}$ 2014), about 160,000 Twitter posts containing "bitcoin” and a positive, negative or uncertainty related term were collected and further analyzed. For instance, the terms "happy", "love", "fun", "good", "bad", "sad" and "unhappy" represent positive and negative emotional signals, while "hope", "fear" and "worry" are considered as indicators of uncertainty. The static (daily) Pearson correlation results show a significant positive correlation between emotional tweets and the close price, trading volume and intraday price spread of Bitcoin. However, a dynamic Granger causality analysis does not confirm a statistically significant causal effect of emotional Tweets on Bitcoin market values. To the contrary, the analyzed data shows that a higher Bitcoin trading volume Granger causes more signals of uncertainty within a 24 to 72-hour timeframe. This result leads to the interpretation that emotional sentiments rather mirror the market than that they make it predictable. Finally, the conclusion of this paper is that the microblogging platform Twitter is Bitcoin’s virtual trading floor, emotionally reflecting its trading dynamics.
\end{abstract}

\begin{document}
\renewcommand{\baselinestretch}{1.25}\normalsize

\maketitle

\section{Abstract \& Keywords}
This paper analyzes correlations and causalities between Bitcoin market indicators and Twitter posts containing emotional signals on Bitcoin. Within a timeframe of 104 days  (November 23$^{rd}$ 2013 - March 7$^{th}$ 2014), about 160,000 Twitter posts containing "bitcoin" and a positive, negative or uncertainty related term were collected and further analyzed. For instance, the terms "happy", "love", "fun", "good", "bad", "sad" and "unhappy" represent positive and negative emotional signals, while "hope", "fear" and "worry" are considered as indicators of uncertainty. The static (daily) Pearson correlation results show a significant positive correlation between emotional tweets and the close price, trading volume and intraday price spread of Bitcoin. However, a dynamic Granger causality analysis does not confirm a causal effect of emotional Tweets on Bitcoin market values. To the contrary, the analyzed data shows that a higher Bitcoin trading volume Granger causes more signals of uncertainty within a 24 to 72-hour timeframe. This result leads to the interpretation that emotional sentiments rather mirror the market than that they make it predictable. Finally, the conclusion of this paper is that the microblogging platform Twitter is Bitcoin’s virtual trading floor, emotionally reflecting its trading dynamics.\footnote{The authors likes to thank Peter Gloor for his feedback on an early version of this paper.}
\\\\
\textbf{Keywords}: Bitcoin, Twitter, Emotions, Sentiments, Prediction, Market Mirror

\section{Introduction}

Bitcoin is a peer-to-peer electronic cash system \cite{nakamoto-2009} and a leading global open-source cryptocurrency \cite{krolldaveyfelten-2013}. On March 8$^{th}$ 2014, one (BitStamp) Bitcoin equals $\$$632.79. The Bitcoin network uses cryptography to control the creation and transfer of money, while transactions are broadcasted as digitally signed messages to the shared public network, the 'block chain'. Bitcoins can be obtained by mining \footnote{For an explanation of the term 'mining' see \citeN{nyt-2013} or \citeN{krolldaveyfelten-2013} } or in exchange for products, services, or other currencies. According to \citeN{wsj-2013}, the commercial use of Bitcoin still seems comparable small, mainly as a result of high and risky price volatilities. However, there are signals of traction in retail business and elsewhere, where Bitcoins are increasingly accepted in transactions. On March March 31st 2014, the total value of Bitcoin amounts\$12.5 billion,\footnote{http://Bitcoincharts.com, March 31st, 2014.} and steadily increasing, news coverage in the recent months pointed towards the legitimate use of the virtual currency, its taxation and the circumstance it might support the trade of illicit goods and services \cite{theeconomist-2013}. Noteworthy, a recent breakdown of the mayor exchange platform Mt.Gox on February 25$^{th}$ 2013 marked a setback for Bitcoin traders and observers.
\\\\
While behavioral economics suggests that emotions can affect individual behavior and decision-making \cite{akerlofshiller-2009,rickloewenstein-2010}, the microblogging platform Twitter has drawn more and more attention from different disciplines as a laboratory to study large sets of social and economic data. Particularly interesting is the influence of Twitter users and information propagation \cite{ye-2010}. For example, \citeN{antweilerfrank-2004} determine the correlation between activity in internet message boards and stock volatility and trading volume, while similar methods on analyzing web communications are for example applied by \citeN{choudhury-2008}, \citeN{gloor-2009} and \citeN{gilbert-2010}. \citeN{zhang-2011}, \citeN{oh-2011}, \citeN{bollenhuinaxiao-2011}, \citeN{jaimes-2012}, \citeN{sprenger-2013} and \citeN{si-2013} conducted work on analyzing microblogging data in correlation with financial time series, i.e. to predict or model the stock market. In terms of results, for instance, the authors \citeN{bollenhuinaxiao-2011} claim an accuracy of 87.6 \% in predicting the daily up and down changes in the closing values of the Dow Jones Industrial Average (DJIA) by using a Google-Profile of Mood States (GPOMS), covering the emotional dimensions of \textit{"Calm, Alert, Sure, Vital, Kind and Happy"}. The authors \citeN{zhang-2011} conclude that tweets relating to $dollar$ have the highest Granger causality relation with stock market indices and thus qualify social media sentiments as a predictor of financial market movements.
\\\\
To the best of our knowledge, there is currently no research paper applying the known methodology of market prediction through microblogging sentiments to the Bitcoin market. However, analyzing the Bitcoin market seems particularly interesting, as it is a global and decentralized 24-hour trading market, with tweets and other information signals that are both virtually and simultaneously provided.
\newpage

\section{Methodology}

By using an automated online collector with access to the Twitter API, a total of 161,200 tweets from 57,727 unique users has been collected within the timeframe from November 23$^{rd}$ 2013 until March 7$^{th}$ 2014.\footnote{N=104} While queries have been updated every hour and grouped on a daily basis, each tweet record provides a source, the timestamp (GMT+0), and the text content of max. 140 characters. 
According to the collector's database\footnote{http://tweetarchivist.com}, the collected tweets created in sum about 300 million impressions, where impressions are the total number of times that the tweets have been delivered to Twitter streams of users. As such, it might be reasonable that even such a small proportion of captured tweets might entail multiplier effects. In order to fetch the relevant "emotional" tweets, following queries have been used: \textit{"$Bitcoin$ AND $feel$ OR $happy$ OR $great$ OR $love$ OR $awesome$ OR $lucky$ OR $good$ OR $sad$ OR $bad$ OR $upset$ OR $unhappy$ OR $nervous$ -bot"} and \textit{"$Bitcoin$ AND $hope$ OR $fear$ OR $worry$"}. By doing so, tweets that contain the word "$Bitcoin$" and one of the other terms (or more) could be archived. The biggest challenge for analyzing Twitter data and preventing statistical bias is undistorted data. Therefore, some data cleaning has been applied in the next step and (for example) the combinations "$happy$ $birthday$" and "$not$ $bad$" have not been counted for "$happy$" or "$bad$" respectively, just as repetitive bot-created content was filtered out as far as possible. We like to point out that retweets are included in our statistics as they might be considered as a strong signal of emotional like-mindedness among users.
\\\\
As a result, four blocks of information can be pooled:

\begin{enumerate}
\item {\bf sum\_positive\_tweets} ($\sum{}$ 131,117): Sum of tweets on Bitcoin containing \textit{positive} signals ($feel$, $happy$, $great$, $love$, $awesome$, $lucky$, $good$).
\item {\bf sum\_negative\_tweets} ($\sum{}$ 19,179): Sum of tweets on Bitcoin containing \textit{negative} signals ($sad$, $bad$, $upset$, $unhappy$, $nervous$).
\item {\bf sum\_emotions} ($\sum{}$ 150,296): Sum of \textit{positive} and \textit{negative} tweets.
\item {\bf sum\_hopefearworry} ($\sum{}$ 10,222): Sum of tweets on Bitcoin containing signals of \textit{uncertainty} ($hope$, $fear$ or $worry$), cf. \citeN{zhang-2011}.
\end{enumerate}
Further, market data from the four leading Bitcoin indices  was fetched, namely BitStamp, Bitfinex, BTC-e and BTC China. \footnote{Together, BitStamp (34\%), Bitfinex (26\%), BTC-e (16\%) and BTC China (10\%) account for about 86\% of the overall Bitcoin market volume. Cf. http://Bitcoincharts.com/charts/volumepie/ (March 31st, 2014)} For each index, following price data has been considered: \textit{$Open$, $Close$, $High$ (intraday), $Low$ (intraday), $Volume$ (BTC), $Volume$ (currency, \$), $Intraday$$-$$Spread$ (\textit{High-Low})}, overall \textit{$Intraday$$-$$Return$ (\textit{$Open$$-$$Close$})} and \textit{$\delta{}$ $Price$ (\textit{$Close$ $_{day+2}$ - $Close$ $_{day0}$} )}. In the following,  \textit{ Figure 1, 2} and \textit{3} visualize the data collection results.
\begin{figure}[h]
\centering
\includegraphics[width=1.0\textwidth]{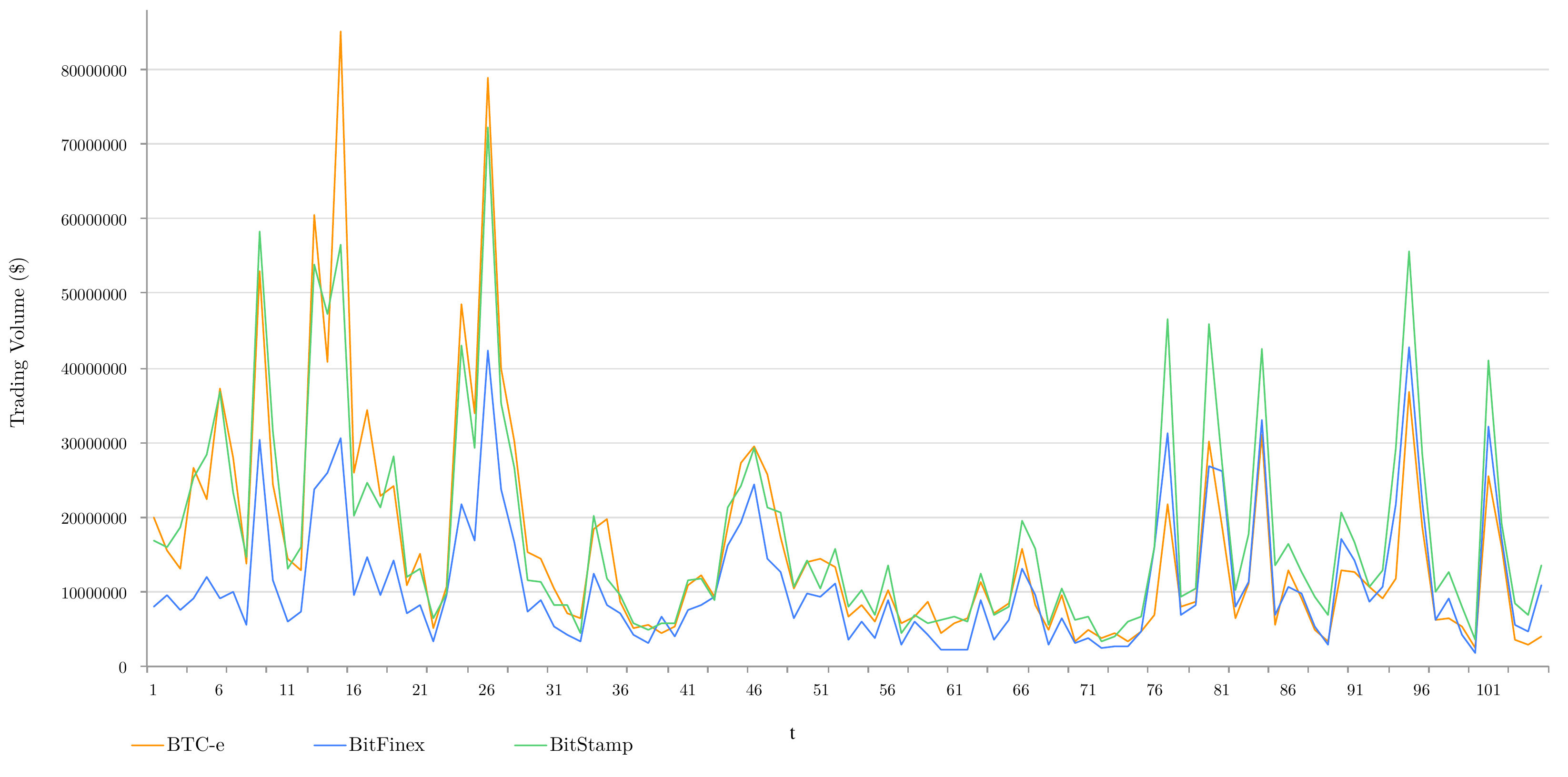}
\caption{Bitcoin markets and their trading volume (by currency, \$). The graph demonstrates the volatility of the Bitcoin market and how fast market shares changed within a period of just $\sim$100 days. For this and all following figures, $x$$-$$axis$ shows $t$ $(days)$.} 
\end{figure}
\begin{figure}[t]
\centering
\includegraphics[width=1.0\textwidth]{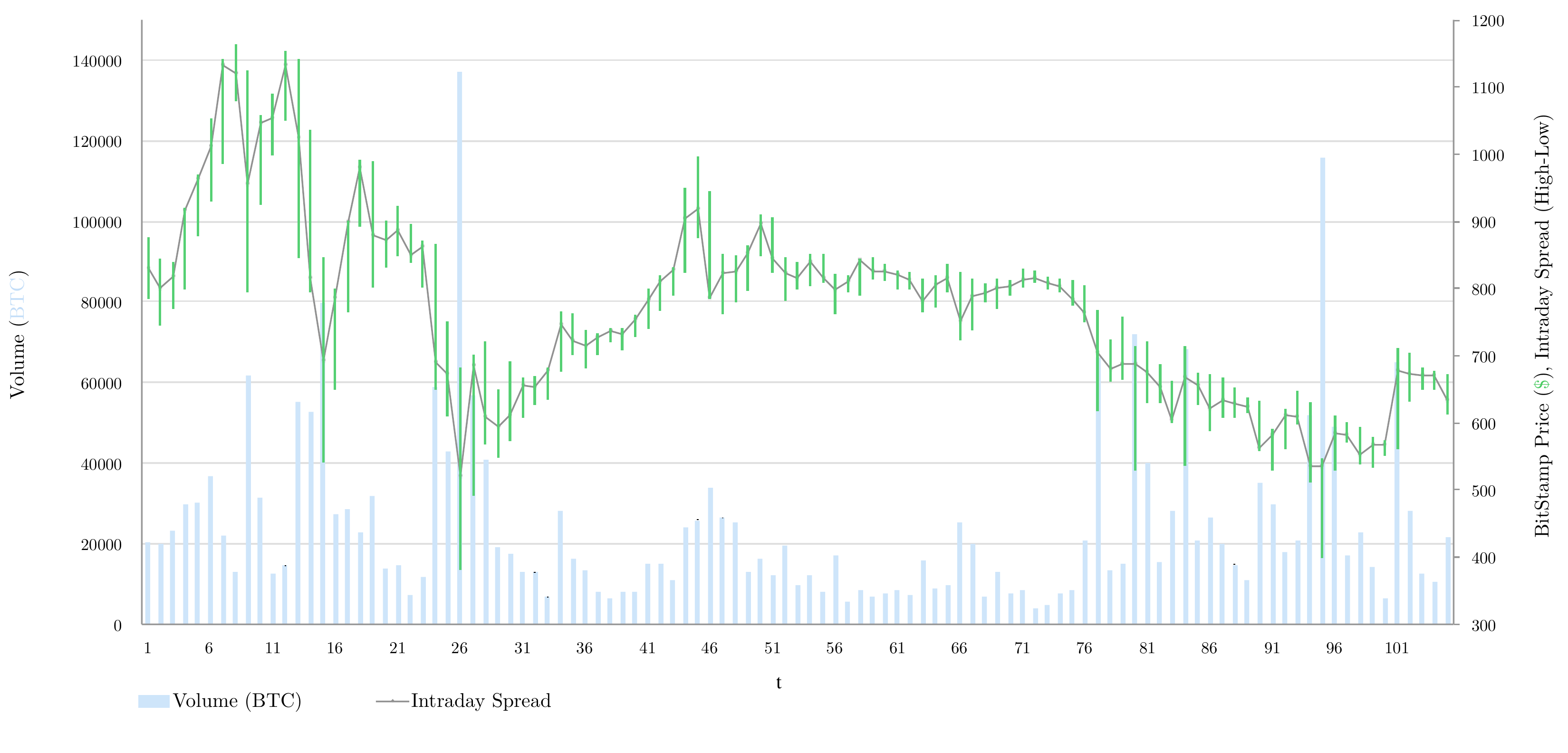}
\caption{BitStamp trading volume ($bitst$$\_volume$) and BitStamp close price ($bitst$$\_close$). Green bars represent the intraday spread (High-Low, $bitst$$\_intraday\_spread$ ). As the graph already suggests, trading volumes are usually higher when the price is lower (significant Pearson correlation of -0.251 at a 0.01 confidence level for $bitst$$\_close$ and $bitst$$\_volume$).}
\end{figure}
\newpage
\begin{figure}[h]
\centering
\includegraphics[width=1.0\textwidth]{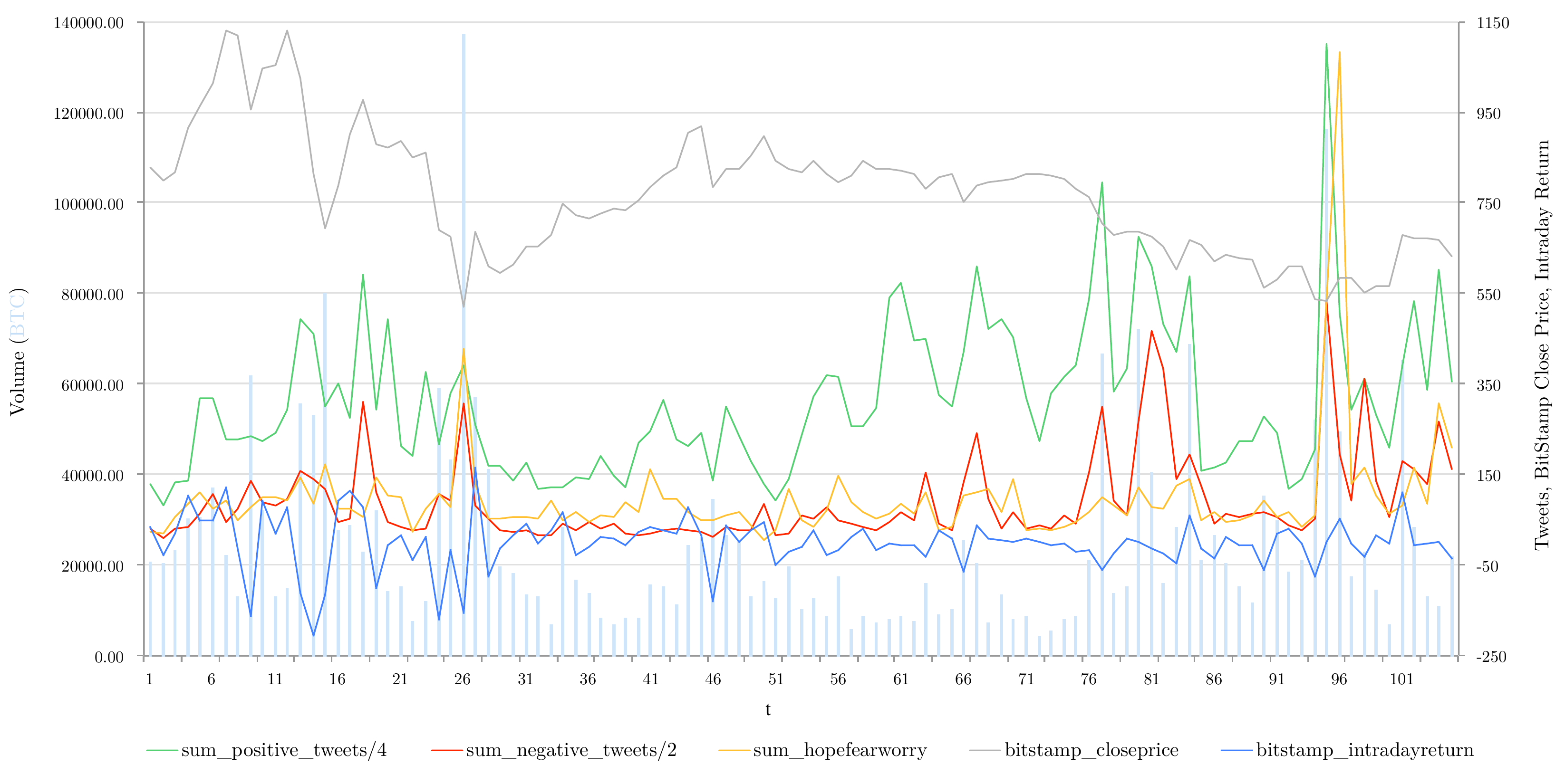}
\caption{Twitter emotions and BitStamp price indicators. $sum$$\_positive$$\_tweets$, $sum$$\_negative$$\_tweets$ and $sum$$\_hopefearworry$ are adjusted (\textit{/x}) to fit the dimensions of the graph. The prominent peaks on day 94-97 can be explained as reactions on Mt.Gox breakdown (February 25$^{th}$ 2014).\\}
\end{figure}

\section{Correlation Analysis}

\begin{table}[h]
\centering
\caption{\bf \bf Bivariate Correlation of Twitter Sentiments and Bitcoin Markets}
\vspace{1em}
\small
\ra{1.1}

\begin{tabular}{l P{2.7cm}P{2.7cm}P{2.7cm}P{2.7cm}}\toprule
& $bitstamp\_closeprice$ & $bitfinex\_closeprice$ & $btce\_closeprice$ & $btcn_\_closeprice$ \\ 
\midrule
$sum\_positive\_tweets$ & -.085 & -.093 & -.054 & -.068 \\
$sum\_negative\_tweets$ & \textbf{-.262**} & \textbf{-.261**} & \textbf{-.278**} & \textbf{.230**} \\
$sum\_emotions$ & -.131 & -.138 & -.109 & -.110  \\
$sum\_hopefearworry$ & \textbf{-.275**} & \textbf{-.249**} & \textbf{-.271**} & \textbf{-.259**} \\
$ratio\_positive\_to\_negative$ & \textbf{.227*} & \textbf{.228*} & \textbf{.282**} & .170 \\
\bottomrule

\end{tabular}
\vspace{0.2em}
\begin{flushleft}
\footnotesize
** \hspace{0.5em}Correlation is significant at the 0.01 level
\\	
\footnotesize
*\hspace{1em} Correlation is significant at the 0.05 level\\
\end{flushleft}
\end{table}
\normalsize
\newpage
$Table$ $I$ shows that the biggest Bitcoin exchange platforms in terms of trading volume seem to be most sensitive for negative tweets and signals of uncertainty ($sum$$\_hopefearworry$) on Bitcoin. Also, a day with a low amount of negative tweets correlates with a higher close price. It can complementary be noted that (in three out of 4 cases) a higher ratio of positive to negative tweets (=more positive than negative tweets) is accompanied by a higher close price. \footnote{It may be noted that the number of positive tweets is much higher than that of negative ones, more than 10 times higher on average. So far, the assumption by \cite{zhang-2011} that people prefer optimistic to pessimistic words can be confirmed.} To the contrary, $positive$$\_$$tweets$ and $sum$$\_$$emotions$ alone do not seem to correlate with the close price. Now, the previously introduced sentiment signals will sequentially be tested with further market indicators for BitStamp, the current biggest exchange market:\footnote{by March 31$^{st}$ 2014}

\begin{table}[h]
\centering
\caption{\bf \bf Bivariate Correlation of Twitter Sentiments and BitStamp Market Indicators}
\vspace{1em}
\small
\ra{1.1}

\begin{tabular}{l P{2.7cm}P{2.7cm}P{2.7cm}P{2.7cm}}\toprule
& $bitst\_closeprice$ & $bitst\_volumebtc$ & $bitst\_intraday$ \newline $\_spread$ & $bitst\_intraday$ \newline $\_return$ \\ 
\midrule
$sum\_positive\_tweets$ & -.085 & \textbf{.393**} & .127 & -.064 \\
$sum\_negative\_tweets$ & \textbf{-.262**} & \textbf{.566**} & \textbf{.286**} & -.151 \\
$sum\_emotions$ & -.131 & \textbf{.452**} & .171 & -.088  \\
$sum\_hopefearworry$ & \textbf{-.257**} & \textbf{.459**} & \textbf{.194*} & -.033 \\
$ratio\_positive\_to\_negative$ & \textbf{.227*} & \textbf{-.426**} & \textbf{-.339**} & .124 \\
\bottomrule

\end{tabular}
\vspace{0.2em}
\begin{flushleft}
\footnotesize
** \hspace{0.5em}Correlation is significant at the 0.01 level
\\	
\footnotesize
*\hspace{1em} Correlation is significant at the 0.05 level\\
\end{flushleft}
\end{table}
\normalsize

\begin{figure}[h]
\centering
\includegraphics[width=1.0\textwidth]{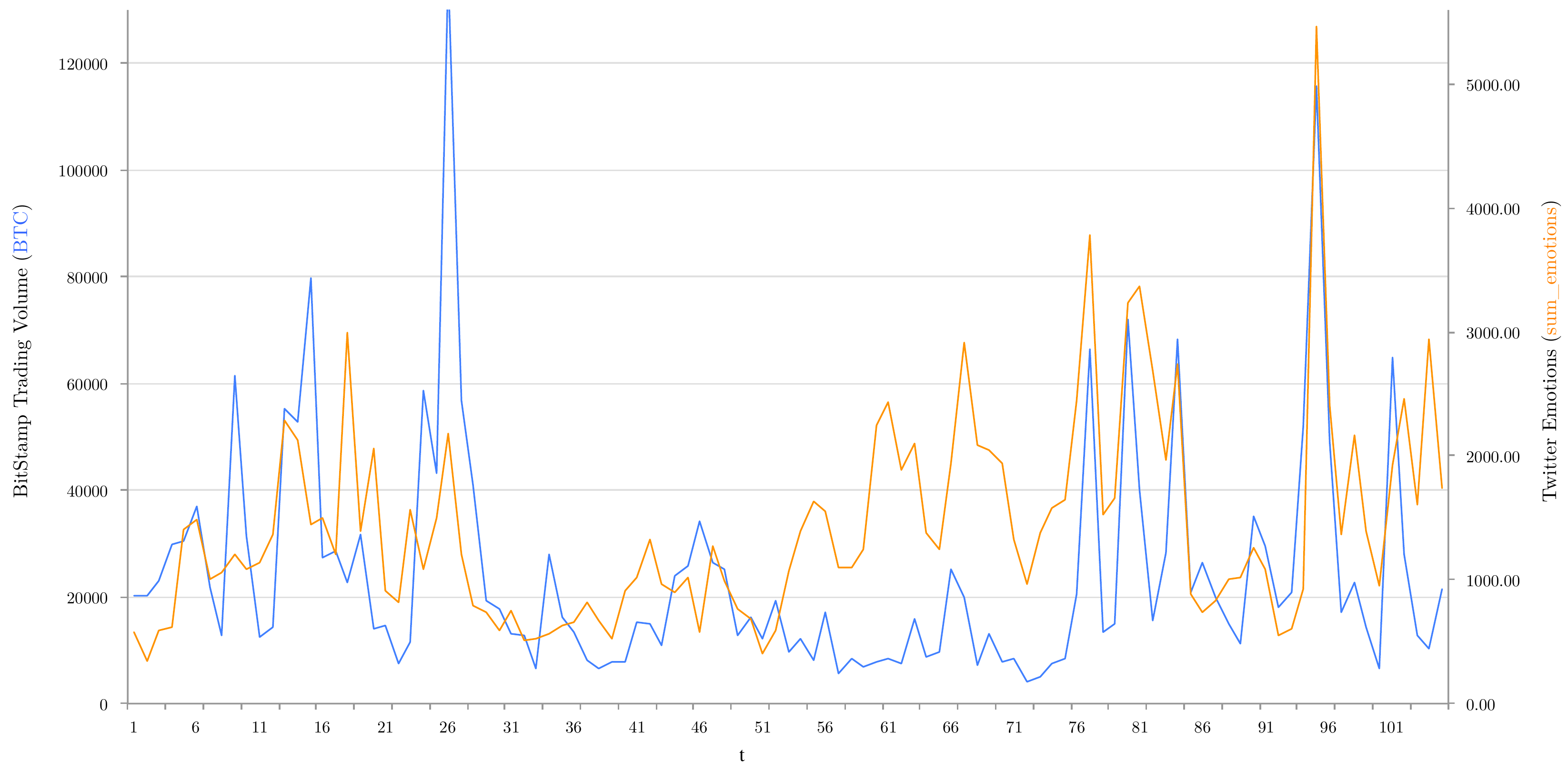}
\caption{BitStamp trading volume ($bitst\_volumebtc$) and emotions on Twitter ($sum$$\_$$emotions$)} 
\end{figure}

Following results can be summarized from $Table$ $II$:
\begin{enumerate}
\item Emotions on Twitter, especially $sum$$\_$$negative$$\_$$tweets$ and $sum$$\_$$hopefearworry$, positively correlate with the BitStamp trading volume (cf. $Fig.$ $4 $ $\&$ $ 5$).
\item The sum of emotions and signals of uncertainty also fuel intraday price volatilities, such as the ($intraday$$\_$$spread$), reflecting the difference between the highest and lowest intraday trading price of BitStamp on a given day.
\item The more negative emotions and signals of uncertainty appear on a given trading day, the more likely is a lower close price. Mixed with signals of uncertainty, negative sentiments may be interpreted as a sign of dissatisfaction or pessimism by traders (and their observers). As $Fig. 5$ illustrates, negative emotions especially seem to appear on trading days with a decreasing close price.
\\
\end{enumerate}

\begin{figure}[h]
\centering
\includegraphics[width=1.0\textwidth]{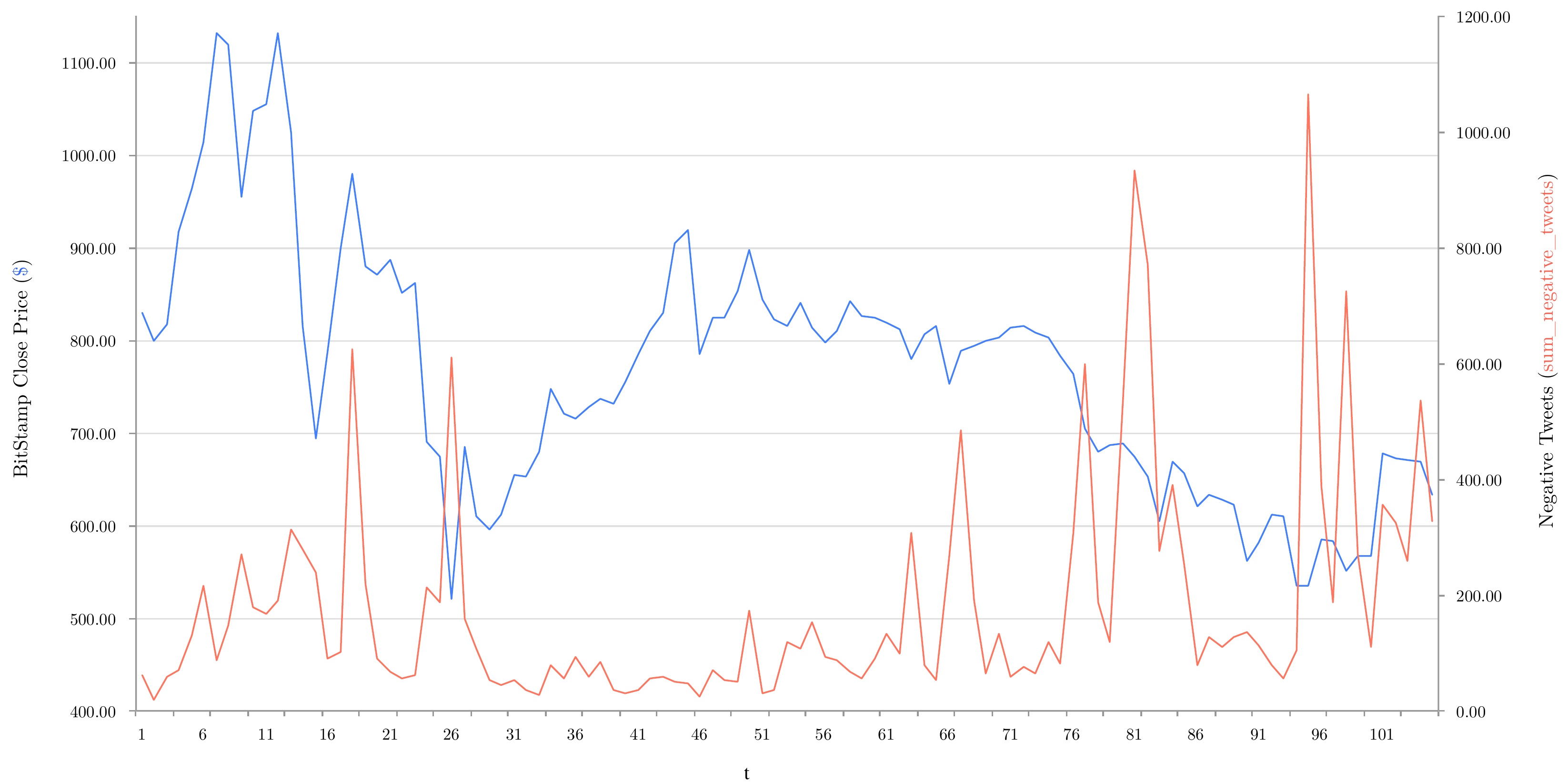}
\caption{BitStamp close price ($bitst\_close$) and negative signals on Twitter ($sum$$\_$$negative$$\_tweets$)} 
\end{figure}

\clearpage
\newpage

\begin{figure}[h]
\centering
\includegraphics[width=1.0\textwidth]{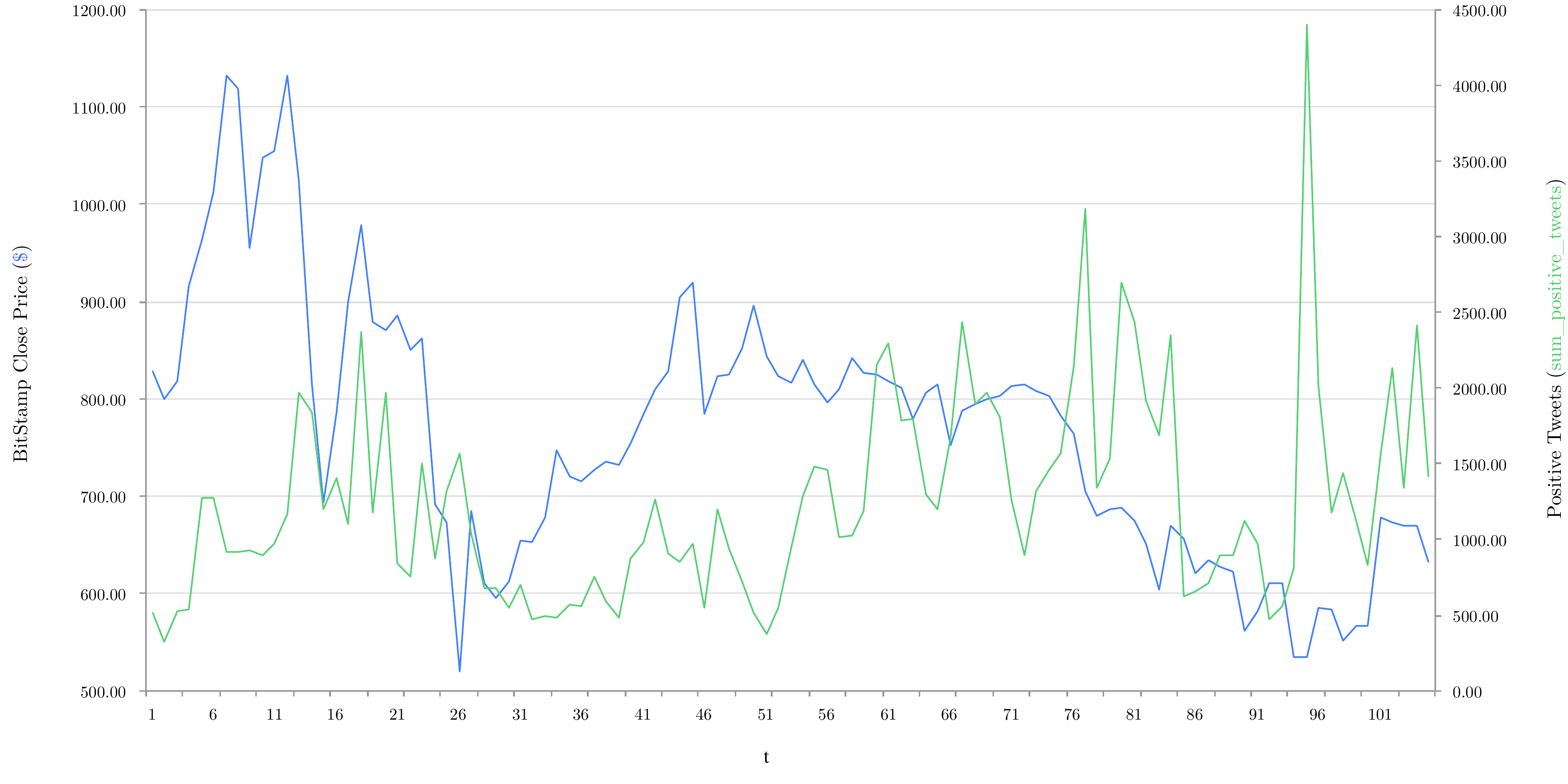}
\caption{BitStamp close price ($bitst\_close$) and positive signals on Twitter ($sum$$\_$$positive$$\_tweets$)} 
\end{figure}

\noindent Current data suggests that the term of 'Nowcasting' (predicting the present), \cite{giannone-2008,choivarian-2012} might be applicable to Bitcoin intraday market development. However, the specific cause and effect relationship in terms of predictability will be tested in $section$ $5$. So far, correlations in this analysis follow the simple and assumption that the relation between variables is linear, which is hardly satisfied in the often appearing random walk of financial market movements. In order to find out more about the prediction value of Twitter sentiments for the development of the BitStamp market, another calculation will be conducted. For this purpose, we will project the BitStamp close price each -2 and +2 days into the past and future respectively, while applying a moving average for related sentiment data.
\begin{table}[h]
\centering
\caption{\bf \bf Bivariate Correlation of Twitter Sentiments and BitStamp Close Price ($\pm{}$ 2 days)}
\vspace{1em}
\small
\ra{1.1}

\begin{tabular}{l P{2.05cm}P{2.05cm}P{2.05cm}P{2.05cm}P{2.05cm}}\toprule
& $-2d\_bitst\_close$ & $-1d\_bitst\_close$ & $bitst\_close$ & $+1d\_bitst\_close$ & $+2d\_bitst\_close$ \\ \midrule
$sum\_positive\_tweets$ & -.022 & -.083 & -.130 & -.175 & \textbf{-.208*} \\
$sum\_negative\_tweets$ & \textbf{-.208*} & \textbf{-.269**} & \textbf{-.318**} & \textbf{-.341**} & \textbf{-.338**} \\
$sum\_emotions$ & -.066 & -.130 & -.180 & \textbf{-.222*} & \textbf{-.248*} \\
$sum\_hopefearworry$ & \textbf{-.262**} & \textbf{-.298**} & \textbf{-.274**} & \textbf{-.256**} & \textbf{-.269**} \\                  
$ratio\_positive\_to\_negative$ & \textbf{.199*} & \textbf{.229*} & \textbf{.263**} & \textbf{-.281**} & \textbf{-.280**}\\
\bottomrule

\end{tabular}
\vspace{0.2em}
\begin{flushleft}
\footnotesize
** \hspace{0.5em}Correlation is significant at the 0.01 level
\\	
\footnotesize
*\hspace{1em} Correlation is significant at the 0.05 level\\
\end{flushleft}
\end{table}
\normalsize
\\\\
Following $Table$ $III$ shows that emotions negatively correlate with the future close price. Especially negative tweets are negatively correlated with the BitStamp close price within a 48-hour timespan (cf. $Fig.$ $5$). Further, the sum of emotions and positive tweets both show a negative correlation with regard to the future close price. Further data suggests that signals of uncertainty (hope, fear, worry) do not only amplify trading volumes (cf. $Table$ $IV$) but also the close price. The higher the amount of uncertainty signals, the lower the BitStamp close price within $\pm{}$ 2days. However, we can assume that tweets are not always "$\grave{a}$ $point$" as timestamps might suggest and thus a time lag in retweets may make an interpretation less robust.

\begin{table}[h]
\centering
\caption{\bf Bivariate Correlation of Twitter Sentiments and BitStamp Trading Volume, Price $\delta{}$ ($\pm{}$ 2 days)}
\vspace{1em}
\small
\ra{1.1}

\begin{tabular}{l P{1.6cm}P{1.6cm}P{1.6cm}P{1.6cm}P{1.6cm}P{1.75cm}}\toprule
& $-2d\_volume$ & $-1d\_volume$ & $+1d\_volume$ & $+2d\_volume$ & $-2d\_\delta{}\_price$ & $+2d\_\delta{}\_price$\\ \midrule
$sum\_positive\_tweets$ & .010 & \textbf{.302**} & .002 & -.045 & -.174 & -.105\\
$sum\_negative\_tweets$ & \textbf{.237*} & \textbf{.505**} & .115 & .107 & -.180 & .015\\
$sum\_emotions$ & .063 & \textbf{.364**} & .028 & -.012 & -.183 & -.082\\
$sum\_hopefearworry$ & \textbf{.362**} & \textbf{.546**} & .025 & -.036 & -.021 & -.028\\                  
$ratio\_positive\_to\_negative$ & \textbf{-.366**} & \textbf{-.447**} & \textbf{-.238*} & -.177 & .014 & -.104\\
\bottomrule

\end{tabular}
\vspace{0.2em}
\begin{flushleft}
\footnotesize
** \hspace{0.5em}Correlation is significant at the 0.01 level
\\	
\footnotesize
*\hspace{1em} Correlation is significant at the 0.05 level\\
\end{flushleft}
\end{table}
\normalsize
\newpage

$Table$ $IV$ demonstrates the correlation between the BitStamp (intraday) trading volume as well as the intraday price spread (High - Low) $\delta{}$ within a $\pm{}$48-hour timespan. As far as data enables an interpretation, a high amount of emotions and signals of uncertainty (in the present) correlates with high trading volumes within the last 24 hours. Especially, a high amount of negative signals is a key influencer for trading volume within the past 24 hours. Again, a more balanced (= lower) ratio of positive and negative sentiments also contributes to a higher trading volume, while for intraday price spreads ($\delta{}$), the current data does not support any significant influence by Twitter sentiments.\\

\newpage

\section{Granger-Causality Analysis}
"Correlation does not imply causation" is a long-known phrase in science. Thus, in order to go beyond correlations and develop a better understanding with regard to the causalities, we apply a Granger causality analysis \cite{granger-1969} to the daily time series of Twitter sentiments and the Bitcoin market movement. Granger causality is a statistical concept of causality that can be used to determine if one time series is useful in forecasting another. A scalar $Y$ is said to \textit{Granger-cause} scalar $X$ if $X$ is better predicted by using the past values of $Y$ than by solely relying on past values of $X$. If $Y$ causes $X$ and $X$ does not cause $Y$, it is said that unidirectional causality exists from $Y$ to $X$. If $Y$ does not cause $X$ and $X$ does not cause $Y$, then $X$ and $Y$ are statistically independent. If $Y$ causes $X$ and $X$ causes $Y$, it is said that feedback exists between $X$ and $Y$.
\\\\
The calculation of Granger causality requires that the time series have to be covariance stationary, so an Augmented Dickey-Fuller \cite{dickey-1979} test has been done first, in which the $H$$_{0}$ ($p$$=$$1$) of non-stationarity was rejected at the 0.05 confidence level for all Twitter and Bitcoin time series variables. All evaluated data is stationary. \footnote{As there was a trend observable for the Bitcoin market development and emotions on Twitter, an analysis with consideration of constant and trend was conducted.} To test whether Twitter emotions “Granger-cause” the changes in the Bitcoin (BitStamp) market, two linear regression models were applied as shown in equations (1) and (2). The first model (1) only uses $p$ lagged values of Bitcoin market data to predict $Y_{t}$ while second model includes the lagged value of Twitter emotions, which are denoted by $X_{t-i}$ . In the given model, we applied a lag $p$ of 1, 2 and 3.

\begin{equation}
	Y_{t} = c_{t}
 + \sum_{i=1}^{p} \beta_{i}  Y_{t-i} + e_{t}
\end{equation}

\begin{equation}
	Y_{t} = c_{t}
 + \sum_{i=1}^{p} \alpha_{i}  Y_{t-i}+ \sum_{i=1}^{p} \beta_{i}  X_{t-i} + u_{t}
\end{equation}

\begin{equation}
with \qquad H_{0} = \beta_{1} = \beta_{2} = ... = \beta_{p} = 0
\end{equation}
\\ After establishing the linear regression equations, $f$ is defined as

\begin{equation}
	f = \frac{\frac{(RSS_{0}-RSS_{1})}{p}}{\frac{RSS_{1}}{(n-2p-1)}} \sim F_{p, n-2p-1}
\end{equation}

where $RSS_{0}$ and $RSS_{1}$ are the two sum of squares residuals of equations $(1)$ and $(2)$ and $T$ is the number of observations.
\\\\ 
If the $F$ statistic is greater than a certain critical value for an $F$ distribution, then we reject the null hypothesis that $Y$ does not Granger-cause $X$, which means $Y$ Granger-causes $X$. As $f$ $\sim F_{p, n-2p-1}$, the question whether $X$ “Granger-causes” $X$ can be solved by checking the value of $p$.

\begin{table}[h]
\centering
\caption{\bf Statistical Significance (p-Value) of Bivariate Granger Causality Correlation Between BitStamp Market Indicators and Twitter Signals.}
\vspace{1em}
\small
\ra{1.1}

\begin{tabular}
{l P{2.7cm}P{2.7cm}P{2.7cm}P{2.7cm}}\toprule
& $bitst\_close$ & $bitst\_volume$ & $bitst\_intraday$ \newline $\_spread$  & $bitst\_intraday$ \newline $\_return$\\ \midrule
$sum\_positive\_tweets$\\
\hspace{0.5em}$Lag = 1$ & (.150) .855 & (.549) .456 & (.591) .755 & (.219) .166\\
\hspace{0.5em}$Lag = 2$ & (.342) .420 & (.225) .027 & (.222) .099 & (.357) .354\\
\hspace{0.5em}$Lag = 3$ & (.434) .499 & (.065) .056 & (.158) .168 & (.436) .510\\
$sum\_negative\_tweets$\\
\hspace{0.5em}$Lag = 1$ & (.309) .322 & (.354) .546 & (.489) .735 & (.677) .860\\
\hspace{0.5em}$Lag = 2$ & (.432) .602 & (.618) .770 & (.419) .907 & (.755) .966\\
\hspace{0.5em}$Lag = 3$ & (.594) .967 & (.174) .941 & (.195) .899 & (.756) .933\\
$sum\_emotions$\\
\hspace{0.5em}$Lag = 1$ & (.158) .721 & (.481) .598 & (.549) .842 & (.272) .251\\
\hspace{0.5em}$Lag = 2$ & (.346) .550 & (.341) .063 & (.233) .185 & (.431) .467\\
\hspace{0.5em}$Lag = 3$ & (.502) .625 & (.064) .139 & (.133) .307 & (.529) .621\\
$sum\_hopefearworry$\\
\hspace{0.5em}$Lag = 1$ & (.434) .091 & (.132) \textbf{.001** $^\alpha$} & (.167) .440 & (.206) .365\\
\hspace{0.5em}$Lag = 2$ & (.249) .196 & (.254) \textbf{.003** $^\beta$} & (.294) .770 & (.247) .189\\
\hspace{0.5em}$Lag = 3$ & (.439) .173 & (.426) \textbf{.009** $^\gamma$} & (.474) .580 & (.489) .098\\
$ratio\_positive\_to\_negative$\\
\hspace{0.5em}$Lag = 1$ & (.504) .410 & (.489) .085 & (.627) .286 & (.859) .593\\
\hspace{0.5em}$Lag = 2$ & (.559) .699 & (.456) .250 & (.960) .632 & (.785) .760\\
\hspace{0.5em}$Lag = 3$ & (.656) .820 & (.654) .381 & (.966) .843 & (.824) .752\\
\\
\bottomrule

\end{tabular}
\vspace{0.2em}
\begin{flushleft}
\footnotesize

$n$ $=$ $102$
\\
Twitter $\rightarrow$ Bitcoin, $X=f(Y)$ in parentheses ; Bitcoin $\rightarrow$ Twitter, $Y=f(X)$ without parentheses.
\\
$\alpha$: $F$ $=$ $12.54$, $\beta$: $F$ $=$ $6.17$, $\gamma$: $F$ $=$ $4.04$
\\
** \hspace{0.5em}Correlation is significant at the 0.01 level
\\	
\footnotesize
*\hspace{1em} Correlation is significant at the 0.05 level\\
\end{flushleft}
\end{table}

\newpage According to $Table$ $V$, there is no significant bivariate Granger causality correlation for $X = f(Y)$ (Twitter $\rightarrow$ Bitcoin) with regard to all market indicators. Twitter have no lagging effect on the Bitcoin market. However, there are bivariate Granger causality correlations for $Y = f(X)$ (Bitcoin $\rightarrow$ Twitter), which means that Bitcoin (BitStamp) market movements induce reactions on Twitter. In particular, $Table$ $V$ indicates that BitStamp trading volume Granger causes signals of uncertainty within a 24 to 72-hour timeframe ($btc$$\_$$volume$ $\rightarrow$ $sum$$\_$$hopefearworry$, cf. $Fig. 7$). This could be interpreted as such that high trading volumes (on average) come along with a high number of transactions. The higher the amount of transactions, the more people on Twitter may articulate their uncertainties by expressing signals of hope, fear or worry.
\\\\
It is noteworthy that Granger causality does not imply "true causality" \cite{granger-2004} and differs from "causation" in the classical philosophical sense. For example, if both $Y$ and $X$ are influenced by a common third variable with different lags, $Y$ might erroneously be believed to Granger-cause $X$.
\\
\begin{figure}[h]
\centering
\includegraphics[width=1.0\textwidth]{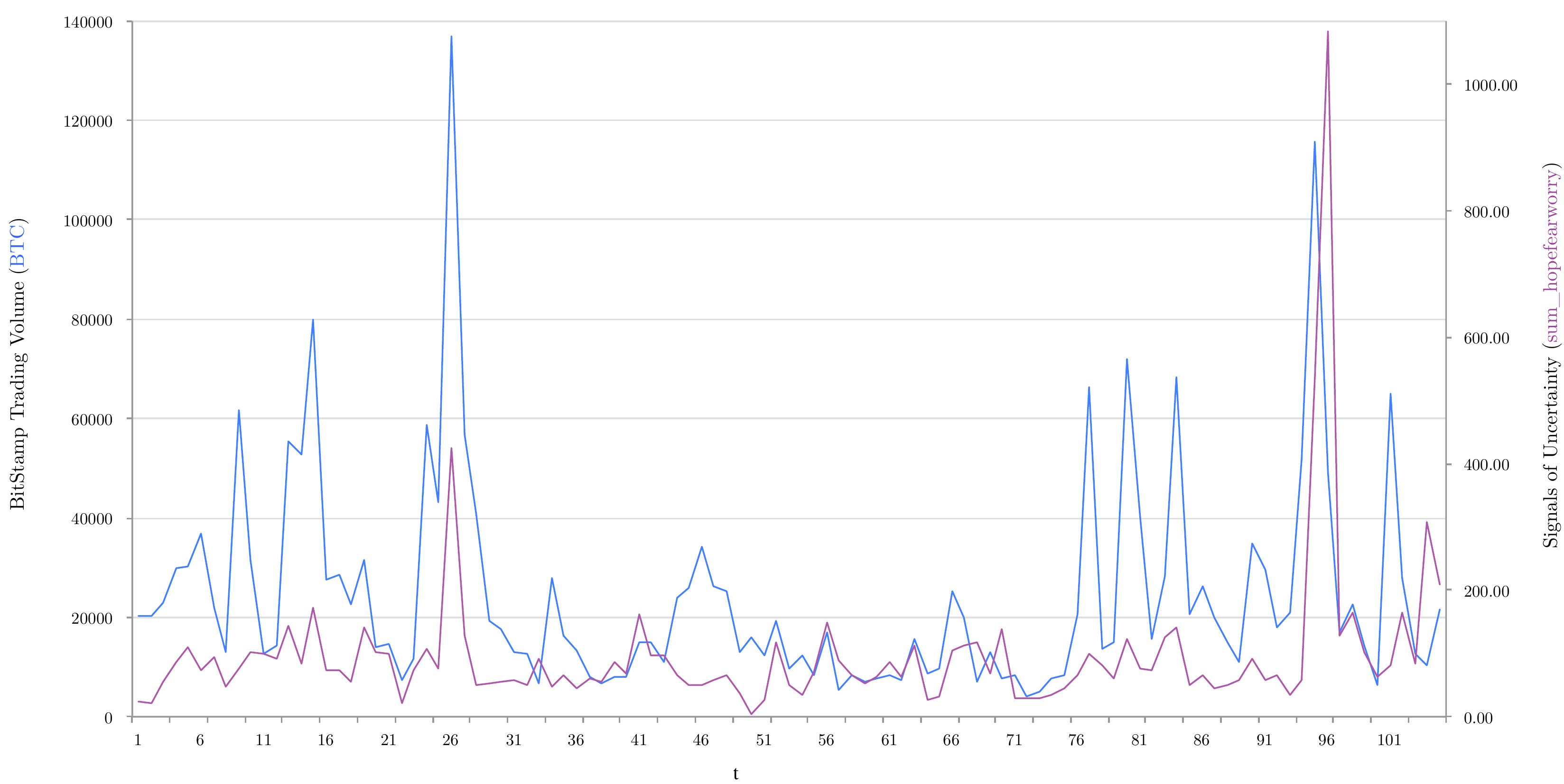}
\caption{BitStamp trading volume ($bitst$$\_volume$) and signals of uncertainty on Twitter ($sum$$\_$$hopefearworry$)} 
\end{figure}
\newpage
\section{Summary}
The main research question of this paper was how far virtual emotions might influence a virtual and decentralized financial market like Bitcoin. Summarizing, we can draw following results:
\begin{enumerate}
\item Static intraday measurements suggest a moderate correlation of Twitter sentiments with Bitcoin close price and volume. Also, a lagged correlation analysis showed that the sum of emotional sentiments and especially negative signals positively correlate with the intraday trading volume within the last 48 hours. This can be translated as follows: \textit{When the trading volume was (and is) high, emotions fly high on Twitter}. As such, Twitter may be interpreted as a place that reflects the "speculative momentum".
\item The Granger causality analysis shows that there is no statistical significance for Twitter signals as a predictor of Bitcoin with regard to the close price, intraday spread or intraday return. To the contrary, results in $Table$ $V$ indicate that the Bitcoin trading volume Granger causes signals of uncertainty within a 24 to 72-hour timeframe. Higher trading volumes Granger cause more signals of uncertainty.
\item
Summarizing the results, the microblogging platform Twitter may be interpreted as a \textbf{virtual trading floor that emotionally reflects Bitcoin's market movement}. 
\end{enumerate}
Keeping up that picture, the imagination of classic open-outcry trading floors comes to mind, where traders shouted and made use of hand signals on the pit. Measuring sound noise on stock exchanges, \citeN{coval-2001} suggests that the communication and processing of highly subtle and complex non-transaction signals (noise) by traders in such an environment plays a central role in determining equilibrium supply and demand conditions. The authors further conclude in 2001 that while trading volumes migrate to electronic exchanges, information from face-to-face interaction might be lost. Now, 13 years after their publication, we might conclude: Maybe, the noise on trading floors is back; just in a different form and space.
\\\\
While the current data of only 104 days already looks promising, a longitudinal analysis of about 6 months might provide a better quality of scientific expressiveness, especially in view of the fact that we currently observe a very volatile market with an observation of 1612 tweets per day on average. Particularly, events such as the breakdown of Mt.Gox can be considered as (both internal and external) market shocks that essentially influence the considered data and statistical methods. Equal attention should be paid to data and sentiment quality, which is very limited in our current methodology. While a better linguistic might significantly improve the quality of data, emotional contagion on Twitter should also be considered as another important factor \cite{hu-2013,coviello-2014}. For example, a TwitterRank \cite {weng-2010} telling more about a user's emotional influence and authencity might contribute to better data quality on the weight of nodes in the communication. Notable in this context is a study by \citeN{hernandez-2014} of about 50,000 messages from more than 6,000 users on Twitter with focus on Bitcoin. The researcher's analysis shows a consistent pattern that people interested in Bitcoin are far less likely to emphasize social relations than typical users of the site. Specifically, Bitcoin followers are less likely to mention emotions (beyond family, friends, religion, sex, and) and have significantly less social connection to other users on the site. If this assumption is true, it can hardly be estimated which effect it might entail for this study.
\\\\
Finally, we cordially invite fellow researchers to keep a close eye on Twitter and the Bitcoin market and to improve the outlined approach.
\\\\
\newpage
\bibliographystyle{ci-format}
\bibliography{ci2014-sample-bibfile}

\end{document}